\documentclass[aps,a4paper,preprint,showpacs]{revtex4}
\usepackage{graphicx}

\begin{document}

\title{High activity of He droplets following ionization of systems inside those droplets}
\date{\today}
\author{Nikolai V. Kryzhevoi}
\email[e-mail: ]{nikolai.kryzhevoi@pci.uni-heidelberg.de}
\author{Vitali Averbukh}
\author{Lorenz S. Cederbaum}
\affiliation{Theoretical Chemistry, Institute of Physical Chemistry at
Heidelberg University, Im Neuenheimer Feld 229, 69120 Heidelberg, Germany}

\begin{abstract}
Relaxation processes following {\it inner-valence} ionization of a system can be modified dramatically 
by embedding this system in a suitable environment. Surprisingly, such an environment can be even 
composed of helium atoms, the most inert species available. As demonstrated by the examples of Ne and Ca 
atoms embedded in He droplets, a fast relaxation process [Interatomic Coulombic Decay (ICD)] takes place 
merely due to the presence of the He surroundings. We have computed ICD widths for both $^4$He$_N$ 
and $^3$He$_N$ droplets doped with Ne and Ca and discuss the findings in some detail. In the case of Ne,
ICD is by far the dominating relaxation pathway. In Ca, atomic Auger decay is also possible but ICD
becomes a competitive relaxation pathway in the droplets.
\end{abstract}

\maketitle

Helium droplets have been attracting considerable attention due to their unique properties. 
We refer the reader to review papers describing the present state of the art of 
the field of helium droplets, its profound achievements and open questions 
\cite{Barranco,Stienkemeier06}. The broad interest in He droplets derives from the recognition 
that they can easily pick up foreign species. Numerous experimental and theoretical studies 
pursued several objectives such as to gain a better insight into the properties of the
pure He drops themselves as well as to derive valuable information on dopants, taking advantage 
of the He droplets as an ultracold and gentle matrix in high resolution spectroscopies. 

Because of the weak impurity-helium interaction, properties of the impurity are usually little 
affected by the He surroundings. However, even minor changes, e.g. small shifts of various 
transitions of a foreign atom due to embedding in a He droplet, are of crucial importance since 
they can shed additional light on the nature of the impurity-helium interaction and are very 
helpful, for example, in predicting the impurity position in the droplet\cite{Stienkemeier04,Reho,Stienkemeier97}. 
Cases where the He environment modifies markedly the properties of a system embedded in a He 
droplet attract special attention. To our knowledge only a few such cases have been reported 
so far, notably in all of them the droplets were subject to ionization\cite{Janda,Meiwes}. 

Irradiation of doped droplets with femtosecond laser pulses causes a rich palette of charged 
ions as a result of droplet disintegration via Coulomb explosion. Although singly charged He 
atoms appear as well\cite{laser1,laser2} the cause for their appearance did not receive the 
attention it deserves. One possibility of the He ions production is obviously the direct                                                                            
ionization by intense laser pulses. Another way one might think of is much more exciting since it
reveals a high activity of the He environment in processes occurring in doped He droplets. Here, 
the He ions are the outcome of a very efficient relaxation pathway of highly excited states of 
the {\it foreign} systems embedded in the droplets which is operative {\it only} due to the 
presence of the He environment. This relaxation process is Interatomic Coulombic Decay (ICD)
\cite{firstICD,Marburger,Jahnke,NEdimer,NEn,Ohrwall,Fano}.

Generally, relaxation of an inner-valence vacancy in an isolated atom proceeds radiatively  
by emission of a photon, in contrast to a deeper core vacancy which decays by ejecting an 
Auger electron. The situation can change drastically if neighboring atoms are present as 
predicted theoretically\cite{firstICD} and confirmed experimentally\cite{Marburger,Jahnke}.  
The presence of a neighboring atom lowers the double ionization threshold of the system 
compared to that of the isolated atom. If this threshold is now below the energy of the 
inner-valence hole state, a novel radiationless decay channel opens. In the ICD process the 
initial inner-valence vacancy is filled by an outer-valence electron from the same atomic 
site while the energy released is transferred to ionize a {\it neighboring} subunit of the 
system. ICD may have dramatic consequences for a system where it takes place making this system
unstable towards fragmentation due to the occurrence of two spatially separated charges. 

ICD is an {\it interatomic} process in its nature and this makes its properties, in particular 
the decay width $\Gamma_{\rm ICD}$ or equivalently the lifetime $\tau_{\rm ICD}$ of 
the inner-valence hole, strongly environment dependent. In the classical example of the 2$s$
ionization of a Ne atom, the $2p\rightarrow 2s$ relaxation energy of 26.91 eV is insufficient 
for emitting another electron from this atom. The Ne$2s$ hole decays, therefore, radiatively in about
175 ps \cite{Griffin} corresponding to $\Gamma_{\rm rad}$ of 3.8 $\mu$eV. On the other hand, if 
a neighboring Ne atom  is present, the 26.91 eV are sufficient to knock a $2p$ electron out 
of this neighboring Ne atom and to make ICD operative \cite{NEdimer,Jahnke}. In the Ne dimer 
$\tau_{\rm ICD}$ is about 85 fs, more than 2000 times smaller than $\tau_{\rm rad}$! The lifetime 
of the Ne$2s$ vacancy reduces further to only a few femtoseconds if the excited Ne atom is 
surrounded by several neighbors like it is the case in a Ne$_N$ cluster \cite{NEn,Ohrwall}. 
\begin{figure}[t]
\includegraphics[angle=270,width=8.6cm]{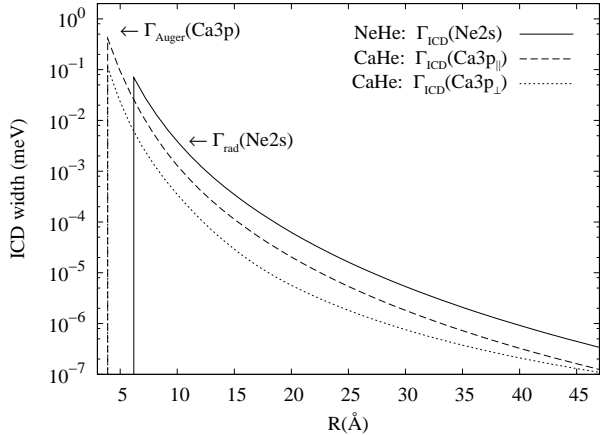}
\caption{ICD widths of inner-valence vacancies in the NeHe and CaHe pairs as functions
of the interatomic distance. The Ca3$p_{\parallel}$ and Ca3$p_{\perp}$ orbitals are parallel 
and perpendicular to the Ca--He axis. The widths of the dominant decay processes in the 
isolated Ne and Ca atoms are labeled.}
\end{figure}

Interestingly, the above mentioned energy of 26.91 eV is also sufficient to ionize a neighboring 
He atom. For this to occur, the minimal separation between He and Ne should be $R_{\rm ICD}=6.2$ \AA\ 
which is derived under the plausible assumption that the ionization energy of He in the presence of the 
Ne ion is governed by electrostatic repulsion between both. Is the presence of a by far more distant He atom
which is less perturbing than Ne able to modify considerably the decay of the Ne2$s$ vacancy like it happens in 
Ne$_2$? To answer this question we have calculated $\Gamma^{\rm pair}_{\rm ICD}(R)$ for a pair of Ne and He 
separated by $R\geq R_{\rm ICD}$ using a recently developed {\it ab initio} Fano-GF method\cite{Fano}. 
The method represents a combination of a Fano formalism for resonance widths, a 
Green's Function method for solving a multielectron problem, and a Stieltjes imaging technique for 
renormalization of the discretized continuum (see Ref.~\cite{Fano} for details). Together with the 
d-aug-cc-pVQZ basis sets for Ne and He \cite{Dunning} we used continuum-like diffuse functions 
of $s$, $p$ and $d$ types\cite{Kaufmann} distributed between these atoms as well as around He. The results 
are shown by the solid curve in Fig.~1. The dominance of ICD over radiative decay at 
$R_{\rm ICD}$ ($\Gamma^{\rm pair}_{\rm ICD}=70 \mu$eV) and some larger $R$ is apparent. 
This fact, being striking by itself, is however of minor consequence for the diatomic NeHe cluster since the 
probability to find Ne and He at such large separations (more than twice the equilibrium distance of the cluster) 
is negligible despite the floppiness of this system.

Being captured by a He droplet, Ne can be surrounded by tens, hundreds or even thousands He atoms,
many of them residing at sufficiently large distances from the impurity to make ICD of the Ne2$s$
vacancy possible. Since the ICD performance increases with opening new ICD channels, ICD is expected to 
be more and more pronounced as the droplet size grows. The big size of a droplet is a serious obstacle 
for calculating ICD widths from first principles. Intuition suggests however that the interaction of 
a dopant with a He atom is not strongly perturbed by other He atoms so that $\Gamma_{\rm ICD}^{\rm drop}$ 
for doped droplets can be reliably approximated by a sum of $\Gamma^{\rm pair}_{\rm ICD}$ for individual 
dopant--He pairs.  The {\it ab initio} calculations below support this view.

Using the Fano-GF method we have explicitly calculated $\Gamma^{\rm clust}_{\rm ICD}$ for the most
symmetric NeHe$_N$ ($N$=1--4) clusters with Ne in the center. The He atoms were separated from Ne by 6.4  \AA,   
a typical radius of the second coordination shell in He droplets doped with Ne (the first 
coordination shell is ICD-inactive).
To keep the size of the problem tractable for the 
{\it ab initio} method used, we had to reduce the quality of the basis set for He to d-aug-cc-pVTZ. 
Excellent agreement between the results obtained and those derived by multiplication of 
$\Gamma^{\rm pair}_{\rm ICD}$ with $N$ (see Fig.~2) strongly suggests the validity of the pairwise 
additivity of ICD widths. Other clusters have also been studied leading to the same conclusions (see
below). 

Assuming pairwise additivity also in larger systems, we can now estimate 
$\Gamma_{\rm ICD}^{\rm drop}$ for Ne@He$_N$ droplets. Ne is known to locate in the 
center of both $^4$He and $^3$He droplets \cite{Scheidemann,Garcias} as a consequence of a stronger Ne--He interaction 
compared to the He--He one. The density distribution $\rho$ of helium  around Ne is spherically 
symmetric and we may write 
$\Gamma_{\rm ICD}^{\rm drop}=4\pi\int_{R_0}^{\infty} \Gamma^{\rm pair}_{\rm ICD}(r) \rho(r) r^2 d\, r$,
where $R_0$ is a cutoff radius below which the He atoms in the droplets are ICD-inactive. 
Due to the combined effect of the He environment, $R_0$ differs slightly from $R_{\rm ICD}$. To assess $R_0$, 
we have calculated ionization thresholds in different NeHe$_N$ clusters, the largest one 
contains 18 He atoms distributed in two shells around the central Ne. The double (Ne$2p$--He1$s$) and 
Ne$2s$ ionization thresholds of these systems were found to be lower than their counterparts in NeHe 
by up to 0.1 and 0.2 eV, respectively, which results in $R_0=5.93$\AA. These 
changes are consistent with the experimentally detected ones of ionization thresholds of He\cite{Denifl} 
and impurities\cite{Loginov} in larger He droplets supporting our results used to determine
the cutoff radius $R_0$ in doped Ne@He$_N$ droplets.  
\begin{figure}[t]
\includegraphics[angle=270,width=8.5cm]{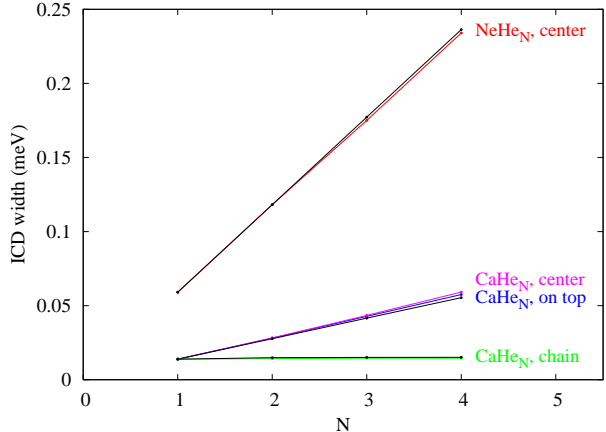}
\caption{(Color online). Comparison of the explicitly {\it ab initio} calculated ICD widths (color lines) 
with those obtained using the pairwise additivity approximation (black lines). Some
nearby curves are hardly distinguishable.}
\end{figure}

We have estimated $\Gamma_{\rm ICD}^{\rm drop}$ for several Ne@$^4$He$_N$ and Ne@$^3$He$_N$ droplets 
(in Fig.~3, circles joined by red and blue lines, respectively) with 50 to 5000 He atoms using 
actual density profiles of the doped droplets $\rho_{\rm dop}$ (filled circles) wherever possible and 
density profiles of the pure droplets $\rho_{\rm pure}$ (open circles) otherwise. All density profiles 
mentioned here and further on have been obtained by Barranco and co-workers within density functional 
theory (see Refs.\cite{Dalfovo,Garcias} for details). Whenever $\rho_{\rm pure}$ had to be used 
(note that this is only the case for $^3$He$_N$) we 
considered $R_0$ as a parameter serving to compensate for using the improper density distribution. Once obtained 
in the case where both $\rho_{\rm dop}$ and $\rho_{\rm pure}$ were available, the new $R_0$ was then 
applied for all droplets with unknown $\rho_{\rm dop}$. $\Gamma_{\rm ICD}^{\rm drop}$ exhibits similar 
behaviors with $N$ for both doped $^4$He$_N$ and $^3$He$_N$ droplets. It rises rapidly for small $N$ approaching a 
saturation limit of 1.8 meV ($\tau_{\rm ICD}$=360 fs) and 1.4 meV ($\tau_{\rm ICD}$=480 fs) at 
$N\sim 1000$ in the doped $^4$He$_N$ and $^3$He$_N$ droplets, respectively.  Astonishingly, 
such drastic reductions of the Ne2$s$ vacancy lifetime to a sub-picosecond timescale compared to that
in the isolated Ne become possible due to the presence and with the active involvement of the He 
environment, the most inert environment available.  
 
The spectacular ability of He environment to modify dramatically the course of processes occurring in 
highly excited species embedded in this environment is obviously not restricted to Ne. Another 
interesting example is provided by doped Ca@He$_N$ droplets. In contrast to the Ne$2s$ vacancy, 
the Ca$3p$ vacancy in an isolated Ca atom can undergo Auger decay. The $4s\rightarrow 3p$ 
relaxation energy of 28.31 eV is more than sufficient for autoionizing an outer-valence Ca$4s$ 
electron resulting in Ca$^{2+}4s^{-2}$. This decay is characterized by a width of 0.58 meV
according to our {\it ab initio} calculation using the Fano-GF method. Radiative decay of the
Ca$3p$ vacancy is much less efficient. The energy of 28.31 eV is high enough for 
ionizing a neighboring He atom via the ICD mechanism.  Irrespective of the direction of the Ca3$p$ orbital, 
the ICD widths in a CaHe pair for all Ca--He separations except for irrelevant short ones are noticeably 
exceeded by $\Gamma_{\rm Auger}$ (see Fig.~1). 

What happens when Ca is ionized in a He droplet?
Auger decay is inherently an {\it intra-atomic} process and its properties are only weakly influenced 
by the very inert He environment. By contrast, ICD is an {\it interatomic} process in character and
its efficiency grows with additional He atoms attached. Thereby, not only the distances from Ca to 
individual He atoms play a crucial role, but also the orientation of the Ca3$p_i$ ($i=x,y,z$) orbital.
Apparently, at a given distance the He atoms along the direction of a particular Ca3$p_i$ orbital 
contribute most to the corresponding ICD width $\Gamma_{i\,\rm ICD}^{\rm drop}$.

Ca is a remarkable element as far as its interaction with He droplets is concerned. Being picked up by a 
$^4$He droplet Ca neither resides in the center nor is attached to surface of the droplet.  It occupies 
an intermediate position in a quite deep surface dimple\cite{Stienkemeier97,Ancilotto}. 
In contrast to this, Ca sits in the center of a $^3$He droplet as predicted recently\cite{Hernando}. 
The various $\Gamma_{i\,\rm ICD}^{\rm drop}$ are equal in Ca@$^3$He$_N$ given that the $^3$He distribution 
is isotropic, and slightly different in Ca@$^4$He$_N$ because the $^4$He distribution is not very 
anisotropic in the close vicinity of Ca residing in a deep dimple. 

Variation of the average ICD width $\bar\Gamma_{\rm ICD}^{\rm drop}=1/3\sum \Gamma^{\rm drop}_{i\,\rm ICD}$ 
with droplet size is of larger practical significance than those of the individual 
$\Gamma^{\rm drop}_{i\,\rm ICD}$. To get insight 
how $\bar\Gamma^{\rm drop}_{\rm ICD}$ changes with the number of He atoms let us first consider three sets of 
small CaHe$_N$ ($N=$1--4) clusters with distinct He arrangements. In the first case Ca is placed 6 \AA\ apart 
from the end of a linear He$_N$ chain with link size of 3.2 \AA. These distances are the typical 
radius of the first coordination shell and the separation between two coordination shells in He droplets doped
with Ca, respectively. Our {\it ab initio} calculations show that all $\Gamma_{i\,\rm ICD}^{\rm clust}$ and 
thus $\bar\Gamma_{\rm ICD}^{\rm clust}$ fulfil pairwise additivity, i.e. in each cluster all explicitly 
calculated $\Gamma_{i\,\rm ICD}^{\rm clust}$ are well reproduced by the sums of ICD widths corresponding to the
available different CaHe pairs (see lower part of Fig.~2 for $\bar\Gamma^{\rm clust}_{\rm ICD}$). 
  
In the other two sets of examples we put the Ca--He separation to 6 \AA\ and arrange helium in such ways to mimic 
solvated and surface locations of Ca. In one case Ca is surrounded by He atoms to form a complex of the highest 
possible symmetry, in the other one Ca is placed on top of the planar, symmetric He$_N$ subcluster with a 
characteristic He-He separation of 3.2 \AA. The following conclusions can be drawn from our calculations.
Although $\Gamma_{i\,\rm ICD}^{\rm clust}$ (not shown) are very sensitive to the arrangement of the He atoms,
$\bar\Gamma^{\rm clust}_{\rm ICD}$ depend only weakly on the He distribution.
The calculated $\bar\Gamma^{\rm clust}_{\rm ICD}$ are shown in the middle 
part of Fig.~2 together with the $N\times\bar\Gamma^{\rm pair}_{\rm ICD}$ values expected when using 
pairwise additivity. In both sets of examples the variations of $\bar\Gamma^{\rm clust}_{\rm ICD}$ with $N$ 
can be well approximated by straight lines whose slopes differ only slightly ($\approx$5\%) from the lines 
$N\times\bar\Gamma^{\rm pair}_{\rm ICD}$. 
\begin{figure}[t]
\includegraphics[angle=270,width=8.5cm]{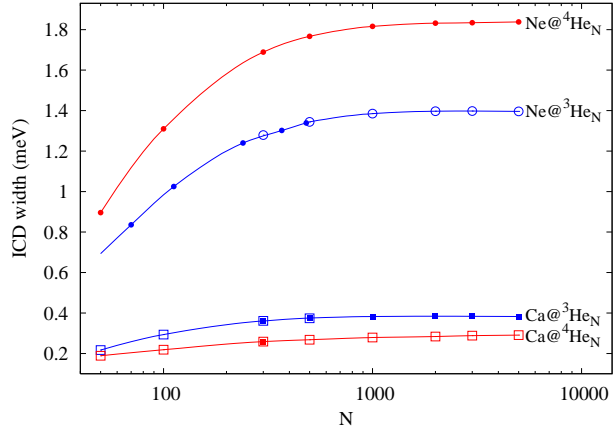}
\caption{(Color online). Estimated ICD widths of inner-valence vacancies in Ne and Ca
doped in $^4$He and $^3$He droplets of different size.  Results obtained by using 
density profiles of doped  and pure He droplets are labeled by filled and open symbols, 
respectively (see text for details). }
\end{figure}

The above findings are of great help for estimating $\bar\Gamma^{\rm drop}_{\rm ICD}$ for large droplets since
the numbers of ICD-active He atoms at each impurity--helium separation are the only quantities which count.
For the case of an axial symmetric $^4$He distribution exhibited in  Ca@$^4$He$_N$ droplets 
we can write $\bar\Gamma_{\rm ICD}^{\rm drop}=2\pi\int_{R_0}^{\infty} \bar\Gamma^{\rm pair}_{\rm ICD}(r)\,r^2\! 
\int_{0}^{\pi}\rho(r,\vartheta) \sin\vartheta\, d\vartheta dr$, where $\vartheta$ is the angle 
between the radius-vector and a line joining the impurity position with the droplet center. 
$\bar\Gamma^{\rm drop}_{\rm ICD}$ for Ca@$^3$He$_N$ droplets which are characterized by a spherically symmetric $^3$He 
distribution can be estimated according to the expression given above for Ne@He$_N$. 
Whenever $\rho_{\rm dop}$ is used we set the cutoff radius $R_0$ to 3.83 \AA\  corresponding to the distance
in the diatomic CaHe at which ICD starts to operate.  Although the actual $R_0$ in droplets is
likely to be slightly different from this value, there is no need to know it precisely because He atoms are 
hardly present at such small separations from the impurity ($\rho_{\rm dop}\approx 0.01\rho_{\rm bulk}$ for 
$r\sim 4.5$\AA\ \cite{Ancilotto,Hernando}, where $\rho_{\rm bulk}$ is the bulk liquid He density). 

The estimated $\bar\Gamma_{\rm ICD}^{\rm drop}$ for both $^3$He$_N$ and $^4$He$_N$ droplets doped with Ca are
shown in Fig.~3 (squares joined by blue and red lines, respectively). The curves exhibit different behaviors 
with droplet size. While $\bar\Gamma_{\rm ICD}^{\rm drop}$ 
in the former case approaches a saturation limit of 0.38 meV ($\tau_{\rm ICD}$=1.7 ps) at
$N\sim 800$,  $\bar\Gamma_{\rm ICD}^{\rm drop}$ in the latter case has not saturated even at  $N=5000$.
Moreover, it shows a clear tendency towards further increase for larger droplets indicating that the nearest
surroundings of Ca (i.e. the dimple and its vicinity) is still under formation there. One sees that the ICD 
width in  doped He droplets has grown markedly compared to that in  CaHe. Although ICD has not become the 
dominant decay process in Ca@He$_N$ it has nevertheless become competitive with  Auger decay 
due to the presence of the He environment.

In summary, unambiguous evidence of the ability of He droplets to modify significantly the course of
ionization processes occurring in systems embedded in them has been furnished. It has been shown that 
the presence of a He environment enclosing Ne and Ca leads to Interatomic Coulombic Decay of their 
inner-valence vacancies. Other suitable dopants including atoms, molecules or clusters can undergo 
the same scenario as well. The pairwise additivity of the ICD width established in small clusters has 
allowed us to estimate the ICD widths in large droplets. The details of 
ICD depend noticeably on droplet size, droplet density and the dopant position and hence also on
the He isotope of which the droplet is made. A comparison of the ICD performance with 
those of competitive relaxation mechanisms in the doped He droplets has revealed that ICD dominates by orders 
of magnitude over radiative decay and matches the performance of Auger decay if the latter is possible. Note 
that ICD might have severe consequences for doped He droplets. The interaction of the He environment
with the slow electrons and ions produced due to ICD can lead to changes of the structure or even to fragmentation 
of the droplets and are issues which merit further theoretical and experimental studies.

We express our gratitude to M. Barranco and his co-workers for valuable discussions and 
for providing us with the He density distributions for the pure and doped He droplets. 
Financial support by the DFG is gratefully acknowledged.

\end{document}